\documentclass{aa}
\usepackage{times,psfig}
\newcommand{\myr}{\ifmmode{{\rm\ M}_\odot{\rm\ yr}^{-1}} \else{${\rm\
         M}_\odot$ yr$^{-1}$}\fi} \newcommand{\msun}{\ifmmode{{\rm\
         M}_\odot}\else{${\rm\ M}_\odot$}\fi}

\begin{document}

   \thesaurus{08 
              (09.16.1; 08.16.4; 09.10.1; 09.11.1; 02.08.1; 03.13.4)}

   \title{Photo-evaporation of clumps in Planetary Nebulae}

   \subtitle{}

   \author{ Garrelt Mellema \inst{1} \and Alejandro C. Raga \inst{2} \and
        Jorge Cant{\'o} \inst{2} \and Peter Lundqvist \inst{1} \and Bruce Balick
        \inst{3} \and Wolfgang Steffen \inst{4} \and Alberto Noriega--Crespo
        \inst{5} } \offprints{G. Mellema}

   \institute{Stockholm Observatory S--133 36 Saltsj{\"o}baden Sweden \and
              Instituto de Astronom{\'\i}a Universidad Nacional Aut{\'o}noma de
              M{\'e}xico, Apdo. Postal 70--264, 04510 M{\'e}xico D.F.  M{\'e}xico \and
              Dept.~of Astronomy, FM--20, University of Washington, Seattle,
              WA 98195, USA \and Dept.~of Physics and Astronomy, University
              of Manchester, Oxford Rd., Manchester M13~9PL, UK \and Infrared
              Processing and Analysis Center, Pasadena, CA~91125, USA }

   \date{Received 2 July 1997; accepted }

   \maketitle

   \begin{abstract} We study the evolution of dense neutral clumps located in
the outer parts of planetary nebulae. These clumps will be photo-ionized by
the ionizing radiation from the central star and change their structure in
the process. The main effect of the ionization process is the setting up of a
photo-evaporation flow and a shock running through the clump. Once this shock
has moved through the entire clump it starts to accelerate because of the
`rocket effect'. This continues until the entire clump has been
photo-ionized. We present an analytic model for the shock and accelerating
phases and also the results of numerical simulations which include detailed
microphysics. We find a good match between the analytic description and the
numerical results and use the numerical results to produce some of the
clump's observational characteristics at different phases of its
evolution. We compare the results with the properties of the fast moving low
ionization knots (ansae or FLIERs) seen in a number of planetary nebulae. We
find that the models match many of the kinematic and emission properties of
FLIERs.
      

   \end{abstract}

%

\section{Introduction}

A fraction of Planetary Nebulae (PNe) show small scale structures which
differ considerably from their surroundings in either line ratios or radial
velocity or both. When these appear in pairs in along the symmetry axis of
the nebulae they are often referred to as `ansae', or when there is a clear
velocity difference, by the acronym of FLIERs (Fast Low Ionization Emission
Regions), first introduced by \cite{Balickea93}. The defining observational
properties of FLIERs are that they should be bright in low ionization lines,
located along the major axis of the PNe, and have a higher velocity than
their environment. There are many cases which fulfill the first two criteria
but for which no kinematic data is available, leaving their status as FLIERs
uncertain.  Examples of these can be found in \cite{Corradiea96}, who took
ratios of H$\alpha$+[N {\sc II}] to [O {\sc III}] to bring out many low
ionization regions in PNe. Often the ansae are only marginally resolved in
ground based images; in some cases they are clearly elongated into
`tails'. In some other cases one observes a whole series of ansae, as for
example in Fg~1 (\cite{Lopezea93}, \cite{Palmerea96}). See \cite{Mellema96}
and \cite{Lopez97} for reviews on what appear to be collimated outflows
from PNe.

Among the best studied FLIERs are the ones in NGC~3242, NGC~6543, NGC~6826,
NGC~7009, and NGC~7662, and these are the ones that have also been selected
for imaging with the {\it HST}\/ (\cite{Harrington95}; \cite{Balickea97}).
Ground based studies of the kinematics of the FLIERs in these PNe can be
found in \cite{BaPrIc87} and \cite{MirSolf92}. The radial velocities of these
FLIERs are typically around 30 to 50~km~s$^{-1}$. \cite{Balickea93} and
\cite{Balickea94}~(1994) studied their spectroscopic properties in some
detail. They found that the derived densities and temperatures are not very
different from those of the environment, a surprising result given their
conspicuous appearance in the low ionization lines. The higher ionization
species are found at the side facing the star, indicating that
photo-ionization determines the ionization stratification. Because of the
observed velocities one would expect a bow shock (and higher ionization
species associated with it) at the end facing away from the star, but this is
not observed.

The {\it HST}\/ results\footnote{See also
http://www.astro.washington.edu/balick/W\_F\_P\_C\_2} (a preview of which can
be found in \cite{WeinKer97}) show a considerable amount of detail and the
analysis has not yet been completed. The images seem to show little arcs
either pointing towards or away from the star, and more diffuse regions also
pointing both towards and away from the star. 

The actual configuration of the FLIERs is not at all clear. The groundbased
observations suggested fast moving clumps being photo-ionized from behind,
but the morphologies seen in the {\it HST}\/ images do not seem to support
this. The tails pointing away from the stars (very clearly seen in NGC~3242),
suggest rather the reverse: a clump being hit by a wind from behind. However
there is no other evidence for such a wind. At this point it seems hard to
propose a definite model for the FLIERs.

One of the surprising properties of the FLIERs is their brightness in [N~{\sc
II}].  The [N~{\sc II}]$\lambda 6583$ line is normally as strong as or
stronger than the nearby H$\alpha$ line. This is usually interpreted as
being due to a high nitrogen abundance, which implies that the FLIERs were
formed from stellar material enriched through the CNO-cycle.

To study the structure of FLIERs we developed a hydrodynamics code which
includes the most relevant micro-physical processes, such as radiative
cooling and photo-ionization. But as was stated above it is not clear what to
choose for initial conditions. There appear to be two cases which could be
used as a FLIER model. Both involve a dense clump moving away from the
central star, but in the one case the clump is moving faster than its
environment, and in the other slower. The first case is the `classical' FLIER
model, but the new {\it HST}\/ results support the second model.

What both cases have in common is the gradual photo-ionization of the
clump. We therefore decided to study this effect first, for the moment
neglecting the effect of any velocity difference with the environment. In
fact, the photo-ionization will cause an acceleration of the clump due to the
so called rocket effect (\cite{OortSp55}~1995). If the rocket effect is very
efficient, it might not even be necessary to invoke a wind; the clump would
reach FLIER-like velocities within its life time ($\sim 1000$~years, see
Balick et al.~1987).

We therefore study here the evolution of a dense clump being photo-ionized by
the stellar radiation. In Sect.~2 we present an analytic model for the
shape and evolution of such a clump. In Sect.~3 we describe the numerical
method that we used to compute the models shown in Sect.~4. In Sect.~5 we
compare the numerical results to the analytic ones, and in Sect.~6 we
compare them to the observations. In Sect.~7 we summarize the conclusions of
the paper.


\section{Analytic model}

\subsection{General description}

The evolution of a dense clump being photo-ionized consists of two phases,
the collapse or implosion phase and the cometary phase. These processes have
been studied by many authors, starting with \cite{OortSp55}~(1955). The three
most recent papers are those of \cite{Bertoldi89} on the collapse phase,
\cite{Bertoldi90} on the cometary phase and \cite{LeflLaz95} on both. The
introduction in \cite{Bertoldi89} contains a good review of all the previous
work.

In the collapse phase an ionization front of type D forms, preceded by a
shock wave which travels through the clump, gradually compressing,
accelerating and heating it. At the ionization front, material is ionized and
thrown out in a photo-evaporation flow. Once the shock has moved through the
entire clump, it enters the cometary phase in which the whole clump starts to
accelerate due to the rocket effect caused by the photo-evaporation flow.

\subsection{The collapse phase}

\begin{figure}
\psfig{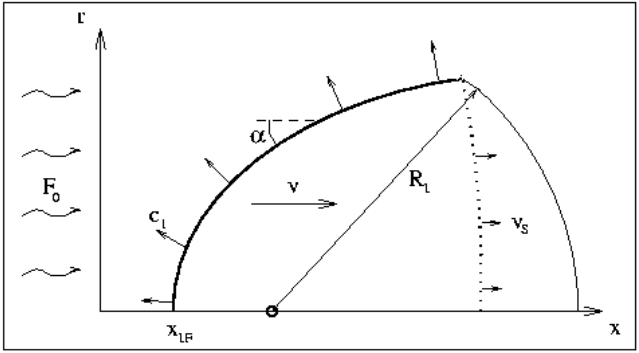}
\vskip 0.1cm
\caption{Schematic diagram showing the interaction of a dense clump of
initial radius $R_{\rm i}$ with an impinging ionizing photon flux $F_0$. The
back-reaction of the photo-evaporation flow (driven out at a sonic velocity
$c_1$) drives a shock (of velocity $v_{\rm s}$) into the clump. The shocked
clump material moves away from the photon source at a velocity $v$. The
position of the ionization front along the symmetry axis is $x_{\rm IF}$. The
ionization front is indicated with a thick line, the shock front with a
dotted line}%
\label{sketch}%
\end{figure}

Let us consider the initial acceleration of a spherical clump by the
interaction with the impinging ionizing radiation field. The clump has an
initial mass $M_{\rm i}$, density $\rho_{\rm i}$, radius $R_{\rm i}$ and
sound speed $c_{\rm i}$. Let $M_{\rm s}$ be the mass of the shocked material
and $v$ its velocity (see Fig.~1). The equation of motion of this mass is
\begin{equation}
M_{\rm s}{dv\over dt}=P_0A_{\rm I}-\rho_{\rm i}v_{\rm s}A_{\rm S}v\,,
\label{collmotion}
\end{equation}
and the equation of mass conservation is
\begin{equation}
{dM_{\rm s}\over dt}=-m\,F\,A_{\rm I}+\rho_{\rm i}v_{\rm s}A_{\rm S}\,,
\label{masscoll}
\end{equation}
where $v_{\rm s}$ is the shock velocity, and $A_{\rm I}$ and $A_{\rm S}$ are
the effective areas of the ionization front and the shock front
(respectively), shown schematically in Fig.~\ref{sketch}. $P_0$ is the
pressure driving the shock, which results from a D-critical ionization
front. $P_0 =2\,F\,m\,c_1$ (see Sect.~2.3), in which $F$ is the ionizing
photon flux reaching the ionization front and $c_1$ is the isothermal sound
speed of the ionized material. We assume $P_0$ to be constant.  The average
mass per atom or ion is taken to be $m\approx 1.3\,m_{\rm H}$ (where $m_{\rm
H}$ is the mass of hydrogen). If $A_{\rm I}\approx A_{\rm S}$,
Eq.~\ref{collmotion} has the trivial solution $v=v_0={\rm constant}$, with
quantities given by
\begin{equation}
P_0=\rho_{\rm i} v_{\rm s} v_0\,,
\label{aeqapr}
\end{equation}
\begin{equation}
\rho_{\rm i} v_{\rm s}=\rho_0\,(v_{\rm s}-v_0)\,,
\label{aeqamom}
\end{equation}
\begin{equation}
\rho_{\rm i}\left({c_{\rm i}}^2+{v_{\rm
s}}^2\right)=\rho_0\,\left[{c_0}^2+(v_{\rm s}-v_0)^2\right]\,,
\label{aeqaen}
\end{equation}
where $\rho_0$ is the density of the shocked material, and $c_0$ its sound
speed. This sound speed is fixed by the radiative losses in the post-shock
region. If we now assume that $c_{\rm i}\ll v_{\rm s}$, Eqs.~\ref{aeqamom}
and \ref{aeqaen} give
\begin{equation}
v_{\rm s}\approx v_0+{{c_0}^2\over v_0}\,,
\label{shockvel}
\end{equation}
\begin{equation}
\rho=\rho_{\rm i}\,\left(1+{{v_0}^2\over {c_0}^2}\right)\,.
\label{shockdens}
\end{equation}
Substituting Eq.~\ref{shockvel} into \ref{aeqapr} one finally obtains the
(constant) velocity $v_0$ of the shocked material, which is given by
\begin{equation}
v_0=\left({P_0\over \rho_{\rm i}}-{c_0}^2\right)^{1/2}\,,
\label{v0}
\end{equation}
so that the position of the clump as a function of time is $x_{\rm
IF}=v_0\,t$. The shock goes through the diameter of the clump in a time
\begin{equation}
t_0={{2\,R_{\rm i}}\over v_{\rm s}}\,,
\label{t0}
\end{equation}
after which the cometary phase starts (described in the next subsection). We
now show that by time $t_0$ a substantial amount of the clump material has
already been photoionized.

Using the $A_{\rm I}\approx A_{\rm S}$ approximation in Eq.~\ref{masscoll},
one can conclude that the ratio between the neutral clump mass at the end of
this phase, $M_0$, and its initial mass is
\begin{equation}
{M_0\over M_{\rm i}}\approx 1-{{m\,F}\over \rho_{\rm i}v_{\rm s}}\,.
\label{shockmass}
\end{equation}

The mass $M$ of the clump can be obtained from the differential equation
\begin{equation}
{dM\over dt}=-m\,F\,A_{\rm I}\,.
\label{massevol}
\end{equation}
In order to integrate this equation, let us consider the length $l$
associated with the area $A_{\rm I}$~: $l\propto {A_{\rm I}}^{1/2}$.  Then,
for a uniform density clump we have $M\propto l^3$, and therefore we can
write
\begin{equation}
A_{\rm I}=\beta\,M^{2/3}\,,
\label{surface}
\end{equation}
where we assume that $\beta$ is constant with time. However, this assumption
of a constant $\beta$ is only strictly true for a clump of time-independent
average density, which is not the case as the clump has both undisturbed and
shocked components (see Fig.~\ref{sketch}) of different densities, and the
proportion of mass in these two components evolves with time. However, as we
will see when comparing the analytic model with the numerical results in
Sect.~5, this approximation appears to give a reasonable description of the
evolution of the clump in the collapse phase.

Using Eq.~\ref{surface} we can integrate \ref{massevol} to obtain
\begin{equation}
M(t)=M_{\rm i}\,\left(1-\beta {m\,F\over 3\,{M_{\rm i}}^{1/3}}t\right)^3\,,
\label{masst}
\end{equation}
where we have taken $M=M_{\rm i}$ for $t=0$ as the initial condition.

To find the constant $\beta$, we use the condition $M=M_0$ (given by
Eq.~\ref{shockmass}) for $t=t_0$. The result is
\begin{equation}
\beta={3\,{M_{\rm i}}^{1/3}\over m\,F\,t_0}\left[1-\left({M_0\over M_{\rm
i}}\right)^{1/3} \right]\,.
\label{beta}
\end{equation}
Substituting Eq.~\ref{beta} into \ref{masst}, we finally obtain the mass of
the clump as a function of time
\begin{equation}
M(t)=M_{\rm i}\,\left\{1-\left[1-\left({M_0\over M_{\rm
i}}\right)^{1/3}\right]{t\over t_0} \right\}^3\,,
\label{msht}
\end{equation}
where the term $M_0/M_{\rm i}$ can of course be substituted by the right hand
side of Eq.~\ref{shockmass}. The flux $F$ during the collapse phase is
approximately given by Eq.~\ref{fapp} (see below) with $R_{\rm IF}=R_{\rm
i}$.

\subsection{The cometary phase}

In this section, we present a description of the `cometary phase', which
begins when the shock has moved through the entire clump. At this time, $t_0$
(Eq.~\ref{t0}), the shocked clump has a mass $M_0$ (Eq.~\ref{shockmass}), and
is moving at a velocity $v_0$ (Eq.~\ref{v0}).

Let us assume that the shocked clump has a uniform temperature $T_0$ (and
therefore a uniform isothermal sound speed $c_0$) and that it is subject to a
uniform acceleration $a$. The pressure and density stratifications then
follow from the condition of hydrostatic equilibrium
\begin{equation}
{dP_0\over dx}=-a\,\rho_0\,,
\label{hydrost}
\end{equation}
which gives pressure and density stratifications of the form
\begin{equation}
P_0(x)=\rho_0(x)\,{c_0}^2=P_0(0)\,{\rm e}^{-x/h}\,,
\label{prstrat}
\end{equation}
where $P_0(0)$ is the pressure at the tip of the clump (at $x=0$), and $h$ is
the scale height, which is given by
\begin{equation}
h={{c_0}^2\over a}\,.
\label{scaleh}
\end{equation}
The $x$-axis points away from the head of the clump in a direction parallel
to the ionizing photon flux (see the schematic diagram of Fig.~\ref{sketch}).

Let us now assume that the clump is embedded in an impinging ionizing
radiation field, so that it is enveloped by an ionization front which moves
into the clump. Let $u_0$ and $u_1$ be the velocities of the neutral and
ionized flows with respect to the ionization front. The mass and momentum
conservation equations then give us
\begin{equation}
\rho_1 u_1=\rho_0\,u_0\,,
\label{masscons}
\end{equation}
\begin{equation}
P_1+\rho_1 {u_1}^2=P_0+\rho_0\,{u_0}^2\,,
\label{momcons}
\end{equation}
where $P_1$ and $\rho_1$ are the pressure and density in the ionized
material, and $P_0$ and $\rho_0$ are the pressure and density on the neutral
side of the front, which are given by Eq.~\ref{prstrat}. From
Eqs. \ref{masscons} and \ref{momcons}, we then obtain
\begin{equation}
P_0(x)=P_1+\rho_1{u_1}^2\left(1-{u_0\over u_1}\right)\,.
\label{prsrel}
\end{equation}
We define $c_1$ to be the isothermal sound speed in the ionized material, and
assume that $u_1=c_1$, as would be the case for a D-critical ionization
front. Then $P_1=\rho_1{u_1}^2$, and therefore
\begin{equation}
{u_0\over u_1}=1-\sqrt{1-\left({c_0\over c_1}\right)^2}\approx {1\over 2}
\left({c_0\over c_1}\right)^2\ll 1\,.
\label{velrat}
\end{equation}
Eq.~\ref{prsrel} can then be written as
\begin{equation}
P_0(x)\approx 2 \rho u_0 c_1\,,
\label{prsrel2}
\end{equation}
where we have also used Eq.~\ref{masscons}.

Let us now consider a point on the surface of the clump, as shown in the
schematic diagram of Fig.~\ref{sketch}. Considering that each ionizing photon
impinging on the ionization front results in a photoionization of a further
atom of the clump, it follows that
\begin{equation}
F\,m\,\sin \alpha \approx \rho\,u_0\,,
\label{ionrate}
\end{equation}
where $F$ is the (uniform) impinging photon flux, $m$ is the average mass per
atom of the clump ($\approx 1.3\,m_{\rm H}$) and $\alpha$ is the angle
defined in Fig.~\ref{sketch}.  In this we ignore the differences in the
ionization structure of hydrogen and helium. Substituting Eq.~\ref{ionrate}
into \ref{prsrel2}, we obtain
\begin{equation}
P_0(x)=2\,F\,c_1\,m\,\sin \alpha\,,
\label{prionrel}
\end{equation}
and from Eq.~\ref{prstrat}
\begin{equation}
P_0(0)=2\,F\,c_1\,m\,,
\label{p0rel}
\end{equation}
leading to the following differential equation for $r$, the cylindrical
radius of the clump at a position $x$,
\begin{equation}
{dr\over dx}=\left({\rm e}^{2x/h}-1\right)^{-1/2}\,.
\label{drdx}
\end{equation}
Integrating Eq.~\ref{drdx} with the boundary condition $r=0$ for $x=0$, one
obtains
\begin{equation}
r=h\,\arctan\left[\sqrt{{\rm e}^{2x/h}-1}\right]\,,
\label{shape}
\end{equation}
which gives the shape of the clump. This solution for the shape of a
photo-evaporating clump is very similar to the solution for a clump
accelerating due to a passing wind (\cite{deyoungaxf67}), normally referred to
as a plasmon. We will therefore also refer to the photo-evaporating clump in
the cometary phase as a plasmon.

The net force ${\cal F}_x$ in the $x$-direction and the mass $M$ of the clump
follow from the integrations
\begin{equation}
{\cal F}_x=\int_0^{\pi h/2} 2\pi r\,P_0(x)\,dr\,,
\label{forceint}
\end{equation}
\begin{equation}
M=\int_0^\infty \pi r^2 \rho_0(x)\,dx\,.
\label{massint}
\end{equation}
Using Eqs.~\ref{prstrat} and \ref{shape}, one can integrate
Eqs.~\ref{forceint} and \ref{massint} to obtain
\begin{equation}
{\cal F}_x=\left(1-{2\over \pi}\right)\,\pi^2P_0(0) h^2\,,
\label{force}
\end{equation}
\begin{equation}
M=\left(1-{2\over \pi}\right)\,\pi^2 {P_0(0)\over {c_0}^2}h^3\,,
\label{mass}
\end{equation}
which enable us to recover Eq.~\ref{scaleh} by taking the ratio $a= {{\cal
F}_x/M}={{c_0}^2/h}$.  Equation~\ref{mass} gives the mass of the clump in
terms of the scale height (or viceversa).

The ionizing photon flux $F$ reaching the ionization front is in general
smaller than the flux $F_0$ delivered by the star at the position of the
clump. The flow produced at the ionization front, streaming out of the clump
towards the star absorbs some of the impinging photons.  The difference
between $F_0$ and $F$ can be expressed as (\cite{Spitzer78}),
\begin{equation}
F_0-F={\alpha_{\rm R} F^2 R_{\rm IF}\over 3\,{c_1}^2}\,,
\label{spitz}
\end{equation}
where $\alpha_{\rm R}$ is the recombination coefficient, and $R_{\rm IF}$ is
the radius of the curved ionization front. Although Eq.~\ref{spitz} has an
exact solution for $F$ in terms of $F_0$ and $R_{\rm IF}$, in order to
proceed analytically with the equations of motion for the clump it is
necessary to use the approximate solution
\begin{equation}
F\approx {F_0\over{\left(1+{\alpha_{\rm R} F_0 R_{\rm IF}\over
3\,{c_1}^2}\right)^{1/2}}}\,,
\label{fapp}
\end{equation}
\noindent which deviates from the exact solution of Eq.~\ref{spitz} by less
than 15\%. For the radius of the ionizing front in Eq.~\ref{fapp}, we will
use the radius of curvature of the tip of the clump, which is equal to the
scale height $h$.

The initial mass $M_0$ and the corresponding scale height $h_0$ are related
through Eq.~\ref{mass}, which gives
\begin{equation}
M_0=\left(1-{2\over \pi}\right){\pi^2\over {c_0}^2}\left(2\,F_0c_1m\right)
{{h_0}^3\over{\left(1+{\alpha_{\rm R} F_0 h_0\over 3\,{c_1}^2}\right)^{1/2}}}\,,
\label{mass0}
\end{equation}
where we have also used Eqs.~\ref{p0rel} and \ref{fapp}.

The rate of change of the mass of the plasmon is related to the impinging
ionizing photon flux through
\begin{equation}
{dM\over dt}=-F\,m\,\pi\,\left({\pi\over 2}h\right)^2\,,
\label{dmdtf}
\end{equation}
where $\pi h/2$ is the outer radius of the projection of the plasmon onto a
plane perpendicular to the symmetry axis (cf.~Eq.~\ref{massevol}). 
Equating Eq.~\ref{dmdtf} to the time derivative of Eq.~\ref{mass} and using 
Eq.~\ref{fapp}, one finds a differential equation for the time evolution
of the scale height $h$
\begin{equation}
{d\tilde{h}\over d\tilde{t}}=-{{6(1+\eta\tilde{h})}\over
{6+5\eta\tilde{h}}}\,,
\label{height}
\end{equation}
in terms of the dimensionless variables $\tilde{h}\equiv h/h_0$ and
$\tilde{t}= t/t_*$, where
\begin{eqnarray}
t_*&\equiv & {v_*h_0\over {c_0}^2}\,,\\ 
v_*&\equiv & {24\over
\pi}\left(1-{2\over \pi}\right)\,c_1\,,\\ 
\eta&\equiv &{\alpha_{\rm
R}F_0h_0\over3\,{c_1}^2}\,.
\label{dim}
\end{eqnarray}
Equation \ref{height} can be integrated with initial conditions $\tilde{h}=1$
for $\tilde{t}=\tilde{t}_0\,(\equiv t_0/t_*)$ to obtain
\begin{equation}
{5\over 6}\left(1-\tilde{h}\right)+{1\over 6\eta}\,\ln\left({{1+\eta}\over
{1+\eta\tilde{h}}}\right)= \tilde{t}-\tilde{t_0}\,.
\label{hint}
\end{equation}

It follows from Eq.~\ref{hint} that the clump becomes fully ionized (that is,
$\tilde{h}=0$) at a time
\begin{equation}
\tilde{t}_{\rm m}=\tilde{t}_0+{5\over 6}+{1\over 6\eta}\,\ln(1+\eta)\,.
\label{tm}
\end{equation}
We find that Eq.~\ref{hint} can be inverted for $\tilde{h}$ in an approximate
way, obtaining
\begin{equation}
\tilde{h}\approx{{\tilde{t}_{\rm m}-\tilde{t}}\over {\tilde{t}_{\rm
m}-\tilde{t}_0}}\,.
\label{happ}
\end{equation}
The mass of the clump as a function of time then follows from Eq.~\ref{mass},
which gives
\begin{equation}
M=M_0\left({{1+\eta}\over {1+\eta\tilde{h}}}\right)^{1/2}\tilde{h}^3\,,
\label{mc}
\end{equation}
with $\tilde{h}$ given by Eq.~\ref{happ}.

The equation of motion of the clump is
\begin{equation}
{dv\over dt}=a\,,
\label{eqmotion}
\end{equation}
where $v$ is the velocity of the clump along the $x$-axis. Using
Eqs.~\ref{scaleh} and \ref{happ}, Eq.~\ref{eqmotion} can be written as
\begin{equation}
{d\tilde{v}\over d\tilde{t}}={1\over \tilde{h}}\,,
\label{eqmotiond}
\end{equation}
where $\tilde{v}\equiv v/v_*$. With $\tilde{h}$ given by Eq.~\ref{happ},
Eq.~\ref{eqmotiond} can be integrated to obtain
\begin{equation}
\tilde{v}=\tilde{v}_0+(\tilde{t}_{\rm
m}-\tilde{t}_0)\,\ln\left({{\tilde{t}_{\rm m}-\tilde{t}_0}
\over{\tilde{t}_{\rm m}-\tilde{t}}}\right)\,.
\label{intvel}
\end{equation}
Note that this expression means that the clump reaches infinite velocity at
$t=t_{\rm m}$, but that this limit is approached logarithmically. The
dimensionless position of the clump $\tilde{x}\equiv x/(v_*t_*)$ follows from
the time integration of Eq.~\ref{intvel}, which gives
\begin{eqnarray}
\tilde{x}=\tilde{v}_0\tilde{t}+(\tilde{t}_{\rm m}-\tilde{t}_0)
\biggl[(\tilde{t}-\tilde{t}_0)+\nonumber\\ \left.(\tilde{t}_{\rm
m}-\tilde{t})\, \ln\left({{\tilde{t}_{\rm m}-\tilde{t}}\over {\tilde{t}_{\rm
m}-\tilde{t}_0}}\right)\right]\,.
\label{intx}
\end{eqnarray}
Taking the limit of Eq.~\ref{intx} for $\tilde{t}\to \tilde{t}_{\rm m}$ we
find that $\tilde{x}\to \tilde{x}_{\rm m}$, where
\begin{equation}
\tilde{x}_{\rm m}=\tilde{v}_0\tilde{t}_{\rm m}+\left(\tilde{t}_{\rm
m}-\tilde{t}_0\right)^2\\,
\label{xm}
\end{equation}
is the position of the clump at the time that it becomes fully ionized. The
velocity averaged over the lifetime of the neutral clump is
\begin{equation}
<\tilde{v}>=\tilde{v}_0+{{\left(\tilde{t}_{\rm m}-\tilde{t}_0\right)}\over
\tilde{t}_{\rm m}}\,.
\label{avvel}
\end{equation}

Equations \ref{v0} and \ref{msht} give the evolution of the clump in the
collapse phase (in which the shock has not yet travelled completely through
the clump). The fully shocked, cometary regime is described by Eqs.~\ref{mc},
\ref{intvel} and \ref{intx}. We find that for the cometary regime it is also
possible to derive analytically the shape of the plasmon (Eq.~\ref{shape}).
In Sect.~5 we carry out a detailed comparison between this analytic
description and a full gasdynamic simulation of the flow.

Our description of the cometary regime is quite similar to the analytic
solution derived by \cite{Bertoldi90}. The model described by these authors
differs from the present one in
the following ways
\begin{itemize}
\item while the present model considers that the incident radiation field has
a planar geometry, the dilution due to a spherical divergence was included by
\cite{Bertoldi90},
\item while \cite{Bertoldi90} consider the effects of the photo-evaporation
flow on the impinging radiative field only in the high absorption limit
(where $F\propto {R_{\rm IF}}^{-1/2}$), our model includes the transition to
the low absorption $F\approx F_0$ regime that occurs for low values of
$R_{\rm IF}$ (see Eqs.~\ref{spitz} and \ref{fapp}).
\item the cases in which magnetic pressure and the thickness of the
ionization front are important were also considered by \cite{Bertoldi90}.
\end{itemize}
The first two points are appropriate for the conditions in PNe, and
necessary to allow quantitative comparisons with the results from the
numerical simulation, which are described in the following sections.


\section{Numerical method}

To numerically study the evolution of a photo-evaporating clump we used the
code described in \cite{RaMelLu97}, adding photo-ionization to the processes
included in the routines dealing with the radiation and ionization
physics. This code (called {\it CORAL}) solves the Euler equations by using a
Van Leer flux-splitting algorithm combined with an adaptive grid method
(\cite{Ragaea95}). The radiation physics part is dealt with using operator
splitting, meaning that it is solved for separately from the hydrodynamics
each time step.

We follow the non-equilibrium evolution of the ionic abundances of H~{\sc
I}--{\sc II}, He~{\sc I}--{\sc III}, C~{\sc II}--{\sc VI}, N~{\sc I}--{\sc
VI}, O~{\sc I}--{\sc VI}, Ne~{\sc I}--{\sc VI}, S~{\sc II}--{\sc VI}. We
assume that C~{\sc I} and S~{\sc I} do not occur due to the abundance of
photons below 13.6~eV. Some of the ionic abundances are derived by
normalizing to the total abundance of the given element (He~{\sc III}, C~{\sc
II}, N~{\sc I}, O~{\sc I}, Ne~{\sc I}, S~{\sc II}). Compared to
\cite{RaMelLu97} we have added nitrogen to the list of elements included
since the nitrogen lines are so prominent in the FLIERs.

The ionic abundances are determined by the recombination rates due to
radiative, dielectronic and charge exchange recombinations, and the
ionization rates due to collisional ionization, charge exchange, and
photo-ionization. For the metals all of these except the photo-ionization
rate are found from two-dimensional tables in temperature and electron
density, calculated using the atomic data specified in \cite{RaMelLu97}.  For
nitrogen we used the atomic data from \cite{LuFr96}, \cite{Nahar95}, and
\cite{Nahar96}. The treatment of the photo-ionization is detailed below.  For
H and He we use analytic fits to the radiative recombination and collisional
ionization rates, and do not consider charge exchange.

From the rates we first calculate the new ionic abundances of hydrogen and
helium, and from these the electron density. This is then used to calculate
the new ionic abundances for the metals iterating a series of two-ionization
state solutions.

Using the updated ionic abundances we calculate the heating rate due to
photo-ionization of H~{\sc I}, and He~{\sc I}--{\sc II}, and the cooling rate
due to line cooling of H~{\sc I}--{\sc II}, C~{\sc II}--{\sc IV}, N~{\sc
I}--{\sc V}, O~{\sc I}--{\sc V}, Ne~{\sc II}--{\sc V}. Similarly to the
ionization and recombination rates, the latter are stored in two-dimensional
tables in temperature and density. The tables for H, C, O, and Ne can be
found in \cite{RaMelLu97}. Tables for nitrogen were compiled from
\cite{LuFr96}, \cite{Bra95}, \cite{PengPradhan95}, and
\cite{Flemingea96}. All tables are available from the authors upon request.
The heating and cooling rates are used to calculate a new temperature. This
whole procedure is iterated several times in order to find the average value
of the temperature during the time step.

\subsection{Photo-ionization processes}

The photo-ionization rates are found from the integrals over the stellar
spectrum weighted with the photo-ionization cross section of the species
under consideration. The heating rates are similar but have the extra factor
of the excess photon energy over the ionization threshold value $\nu_{\rm
t}$.  The treatment of these integrals is similar to the approach described
in \cite{FrMel94}. We basically calculate the values of the integrals for a
range of optical depths for three frequency intervals: $\nu_{\rm t}(\mbox{H~{\sc I}})$--$\nu_{\rm t}(\mbox{He~{\sc I}})$, $\nu_{\rm t}(\mbox{He~{\sc
I}})$--$\nu_{\rm t}(\mbox{He~{\sc II}})$, and $\nu_{\rm t}(\mbox{He~{\sc
II}})$--$\infty$.  In principle one has to deal with two optical depths
($\tau(\mbox{H~{\sc I}})$ and $\tau(\mbox{He~{\sc I}})$) in interval 2, and
three (those and $\tau(\mbox{He~{\sc II}})$) in interval 3. But it is possible
to represent these sums with one optical depth by using the following
approximation taken from \cite{TenBodNor83}:
\begin{eqnarray}
\tau_2(\nu)&=&\bigl[\tau_{\rm t}(\mbox{H~{\sc I}})(0.63\nu_{\rm t}(\mbox{H~{\sc
                  I}}))^{1.7}+\nonumber\\ &&\quad\tau_{\rm t}(\mbox{He~{\sc
                  I}})\nu_{\rm t}^{1.7}(\mbox{He~{\sc I}})\bigr]\nu^{-1.7}
\label{tau2}
\end{eqnarray}
\begin{eqnarray}
\tau_3(\nu)&=&\bigl[\tau_{\rm t}(\mbox{H~{\sc I}})\nu_{\rm t}^{2.8}
     (\mbox{H~{\sc I}})+ \nonumber\\
          &&\quad\tau_{\rm t}(\mbox{He~{\sc I}})(1.51\nu_{\rm t}(\mbox{He~{\sc
             I}}))^{2.8}+\nonumber\\ 
          &&\quad\tau_{\rm t}(\mbox{He~{\sc II}})\nu_{\rm t}^{2.8}
             (\mbox{He~{\sc II}})\bigr]\nu^{-2.8}
\label{tau3}
\end{eqnarray}
in which the $\tau_{\rm t}$ and the $\nu_{\rm t}$ are the values at the
ionization threshold of each interval. These formulae assume that the
cross-sections behave as powerlaws with index 2.8 for H~{\sc I} and He~{\sc
II}. For He~{\sc I} we use an index of 1.7 which is accurate between 24.6~eV
and 45~eV, and at 54.4~eV gives only a $\sim 15$\% overestimate of the cross
section (\cite{Samsonea94}).

Storing the values for the three frequency ranges for 40 optical depths
ranging logarithmically from $10^{-4}$ to $10^4$, we only need to calculate
the integrals once and afterwards interpolate in the tables. By using the
approximation of \cite{TenBodNor83} we correctly describe the hardening of
the spectrum close to an ionization front and the effects of the presence of
He, and thus improve on the approach described in \cite{Ragaea97}, without
having to recalculate the integrals every time step.  The necessary
photo-ionization cross-sections were taken from \cite{Osterbrock89} and
\cite{Cox70}.


\section{Numerical results}

We ran a simulation of the evolution of a clump using the method described
above. As parameters we chose an initial clump density of 5000~cm$^{-3}$,
temperature 100~K, distance to the star $10^{18}$~cm$=0.32$~pc and diameter
$2\times10^{16}$~cm. With these parameters the initial mass of the clump is
$2.3\times 10^{-5}$~M$_\odot$. The clump is initially neutral, except for the
C~{\sc II} and S~{\sc II} (see Sect.~3). The environment has a density of
100~cm$^{-3}$, temperature $10^4$~K, and is initially fully ionized. The star
has a black body spectrum with an effective temperature of $50\,000$~K, and
luminosity 7000~$L_\odot$, typical values for the central stars of young
PNe. For abundances we used the PN values in Table~5.13 from
\cite{Osterbrock89}. The size of the adapative grid at maximum resolution is
$1024\times 256$.

We followed the evolution of the clump for 950~years after which it has been
completely evaporated. We find that the initial collapse phase lasts about
300~years, and that the rest of the time the clump can be described as an
accelerating plasmon. Here we show one snapshot from the collapse phase
($t=250$~years, Fig.~2) and two from the cometary phase ($t=500$ and
750~years, Figs.~3 and 4).  These figures show the density and velocity field
for these times. The lower plot also shows the extent of the clump material
(mainly due to the photo-evaporation) and the areas which are neutral. The
velocity arrows are only plotted if they correspond to velocities higher than
5~km~s$^{-1}$.

\begin{figure*}
\psfig{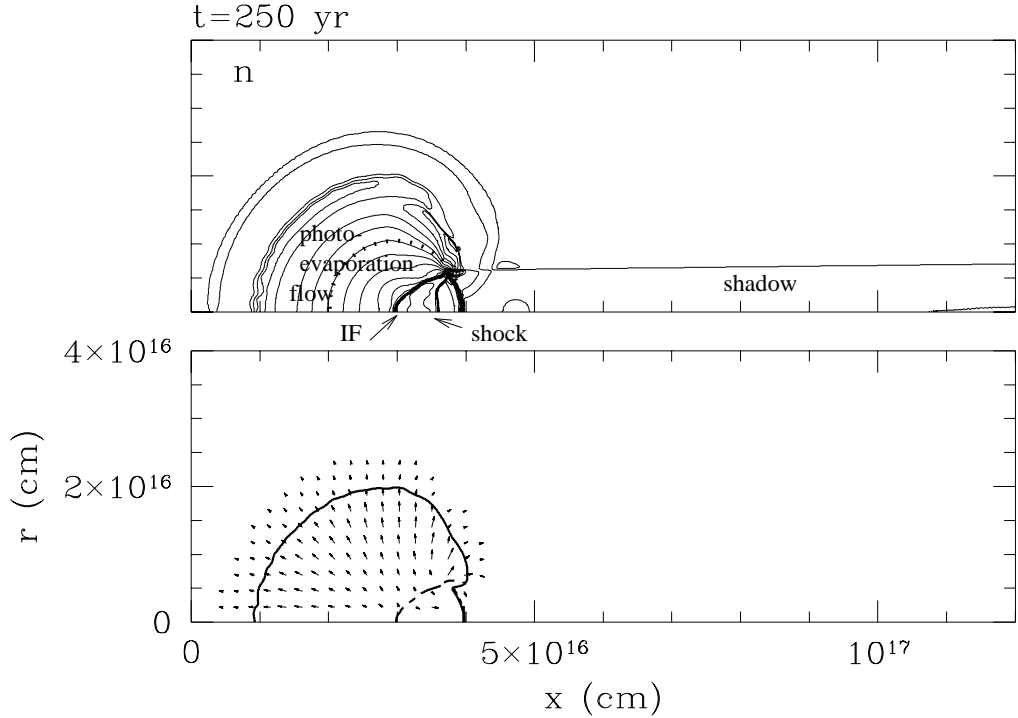}
\vskip 0.1cm
\caption{\protect{Density and velocity at $t=250$~years.
The velocity arrows are only plotted for velocities larger than
5~km~s$^{-1}$; in the velocity plot the solid line deliminates the extent of
the original clump material and the dashed line is the 50\% H ionization
contour. The dotted demicircle in the top figure indicates the original
position and size of the clump. The contours in the density plot are factors
$\sqrt{2}$ apart; the lowest contour is 0.1~cm$^{-3}$}}%
\label{t250}%
\end{figure*}
At $t=250$~years, the clump which originally extended from $x=2\times
10^{16}$ to $x=4\times 10^{16}$~cm, has evaporated about half its
material. The ionization front is at $3\times 10^{16}$~cm and the shock front
at $3.6\times 10^{16}$~cm. One sees the photo-evaporation flow at the back of
the clump; th flow's density drops as it expands and accelerates away from
the clump, and there is a double shock structure separating it from the
environment material. The area in the shadow of the clump is recombining and
cooling, which means that it is no longer in pressure equilibrium with the
rest of the environment material. It is slowly being compressed and heated;
this is more apparent at later times.

The compression through the shockwave is about a factor 5, leading to
densities of about $25\,000$~cm$^{-3}$ in the shocked part of the clump. The
temperature in that part is a little under $10^4$~K, not enough to produce
appreciable collisional ionization, and the material is therefore almost
completely neutral. It is accelerated to about 10~km~s$^{-1}$.

\begin{figure*}
\psfig{figure=t20a.epsi,width=13.5cm,angle=270}
\vskip 0.1cm
\caption{Density and velocity at $t=500$~years. See Fig.~\ref{t250} for more
details}
\label{t500}%
\end{figure*}

At $t=500$~years, the shock wave has travelled through the entire clump. The
shadow region has become more compressed, and there is a small part at the
front of the plasmon which actually has formed from environmental shadow
material. The velocity arrows show that the clump has started to accelerate,
as is also witnessed by the fact that the photo-evaporation flow on the top
of the plasmon leaves it at an angle. In the rest frame of the clump the
evaporation flow is perpendicular to the surface. More details about this
phase can be found in the next section where we compare it to the analytic
solution. The plasmon density ranges from about $2\times 10^4$~cm$^{-3}$ near
the ionization front to 1500~cm$^{-3}$ at the far end; temperatures vary from
$10^4$~K to 2000~K. Velocities vary from 13 to 20~km~s$^{-1}$. Due to the
interaction with the environment material the shock in the photo-evaporation
flow has acquired a conical shape.

\begin{figure*}
\psfig{figure=t25a.epsi,width=13.5cm,angle=270}
\vskip 0.1cm
\caption{Density and velocity at $t=750$~years. See Fig.~\ref{t250} for more
details}
\label{t750}%
\end{figure*}

At $t=750$~years, the clump is close to complete evaporation. The neutral
region of the flow has become much more extended due to the recombination of
the environment material in the shadow of the clump.  Otherwise the situation
resembles the one at $t=500$~years. At the tip the density is still high
($1.5\times10^4$~cm$^{-3}$), but in most of the plasmon it is a factor 10
lower than that. The temperature at the tip is about 6000~K, in the rest of
the plasmon it varies between 2000 and 3000~K. The velocity throughout the
plasmon is about 20~km~s$^{-1}$.

After this the clump quickly becomes thinner and thinner, until it completely
disappears (between 900 and 950 years). The neutral former shadow region
stays on for about 50 years more, after which it becomes photo-ionized again,
and all the material on the grid is ionized. This accounts for the last data
point in Fig.~5.

We find that the photo-evaporation flow has a temperature of $\sim
3000$--5000~K at the base. Within the ionization front, the temperature has a
sharp peak of more than $10^4$~K, a result of the hardening of the spectrum
within the front.


\section{Comparison between the analytic and numerical models}

\begin{figure}
\psfig{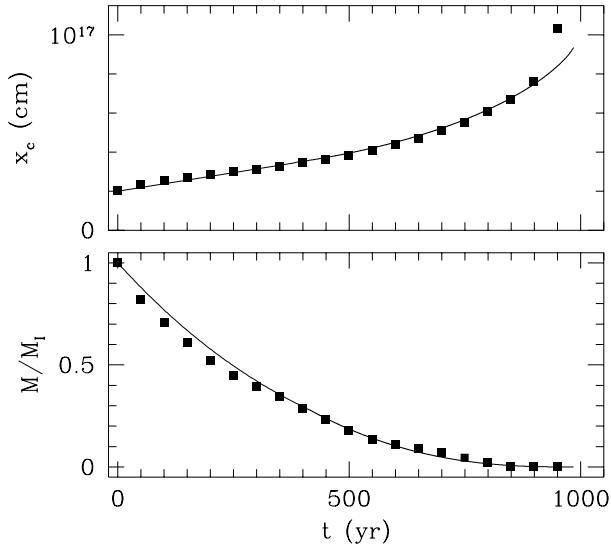}
\vskip 0.1cm
\caption{Position and mass of the clump against time. Squares are values from
the numerical simulation, the solid lines are the results of the analytic
theory}
\label{posmass}%
\end{figure}

To compare the numerical results from the previous section with the analytic
description from Sect.~2, we determined the position and mass of the clump as
a function of time. The position of the clump is defined as the position on
the symmetry axis where $n(\mbox{H~{\sc II}})/n(\mbox{H})=0.5$, and the mass is
all of the clump material with $n(\mbox{H~{\sc II}})/n(\mbox{H})<0.5$. The
squares plotted in Fig~\ref{posmass} are the values we derived. After
900~years there is no neutral clump material left.

In order to carry out a comparison with the analytic model, in Fig.~5 we also
plot the results obtained from Eqs.~\ref{v0} and \ref{intx} (giving the
position as a function of time) and Eqs.~\ref{msht} and \ref{mc} (giving the
mass of the clump as a function of time).

\begin{figure*}
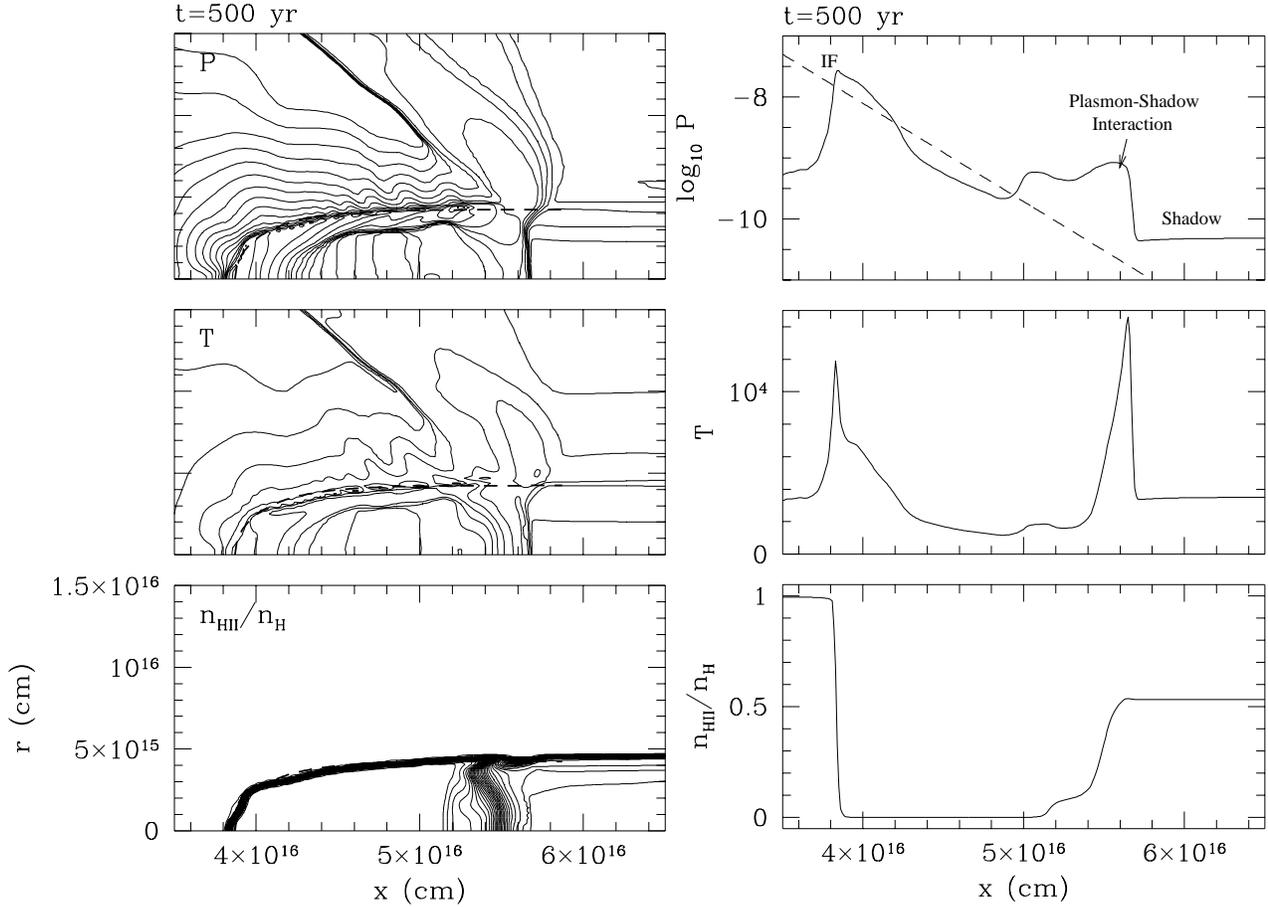

\centerline{\psfig{figure=plasan.epsi,height=12cm}
\psfig{figure=plascuta.epsi,height=12cm,angle=270}}
\vskip 0.1cm
\caption{Left: Contour plots of the pressure, temperature and H ionization
fraction at $t=500$~years. The dashed line shows the fit of Eq.~\ref{shape}
with $h=2.7\times 10^{15}$~cm. The contours in the pressure and temperature
plots are logarithmic with a factor $\sqrt{2}$ between each contour and
minima of $1\times 10^{-10}$~erg~cm$^{-3}$ and 100~K. The contours in the
ionization plot are linear from 0.02 to 0.88 in steps of 0.02. Right: Cuts
along $r=0$ for the three contour plots. The dashed line in the pressure plot
shows the pressure gradient corresponding to the scale height found from
fitting the shape of the plasmon.  }
\label{shapecomp}%
\end{figure*}


The free parameters of the analytic model are the initial density and radius
of the clump (which we set equal to the corresponding values of the numerical
simulation), and the sound speeds $c_0$ (of the shocked clump) and $c_1$ (of
the photo-evaporation flow). As discussed above, from the numerical
simulation we see that the shocked clump has a temperature of $\sim 6000$~K,
and that the base of the photo-evaporation flow has a temperature of $\sim
3000-5000$~K. From these temperatures, we derive values $c_0\approx
6.2$~km~s$^{-1}$ and $c_1\approx 8.4$~km~s$^{-1}$ for the shocked clump and
the ionized flow, respectively. For these parameters, we obtain
$t_0=415$~years, $t_{\rm m}=985$~years, $h_0/R_{\rm i}=0.344$, and an average
velocity $<v>=26$~km~s$^{-1}$ over the lifetime of the neutral clump.

We also compared the shape of the plasmon with the analytic expression given
by Eq.~\ref{shape}. The left-hand side of Fig.~6 shows contour plots for the
pressure, temperature and H ionization fraction at $t=500$~years with a fit
according to Eq.~\ref{shape} using $h=2.7\times 10^{15}$~cm. The steep
gradient seen at $5.5\times 10^{16}$~cm corresponds to the transition from
the clump to the shadow behind it. The fit for the shape is very good,
especially when one considers the complex structure seen inside the
plasmon. To check whether the derived scale height corresponds indeed to the
real one, we have plotted a cut of pressure along the symmetry axis and
overplotted the gradient expected from a value of $h=2.7\times
10^{15}$~cm. This is shown in right-hand side of Fig.~6, together with cuts
for the temperature and H ionization fraction. One sees that logarithmic
pressure gradient is not constant but on average corresponds quite well to
the value used to the make the fit to the shape of the plasmon.

Another method to derive the scale height is to use Eq.~\ref{happ}. Using
$t=500$~years and the values of $t_0$, $t_{\rm m}$ and $h_0$ derived above,
we obtain a value for the scale height $h={2.9\times 10^{15}}$~cm, in
excellent agreement with the value obtained from the fit to the shape of the
plasmon.

Given the very simple nature of the analytic model and the many details
included in the numerical model, the match between the two is surprisingly
good. Both the temporal behaviour (Fig.~5) and the shape of the plasmon
(Fig.~6) found in the numerical simulation closely match the analytic
ones. This can be seen both as a test of the numerical method and a
confirmation that the assumptions going into the analytic models are
largely valid. The match is even more surprising given the poor match found
between the analytic and numerical results for a plasmon being accelerated
through the interaction with a wind (see e.g.\ \cite{Nittmea82}).

   \begin{table*} \caption{Line strenghts (in $L_\odot$)} \label{lines} \[
      \begin{tabular}{llllll} \hline \noalign{\smallskip} Time (yrs) &
      H$\alpha$ & [N~{\sc II}]$\lambda$6583 & [N~{\sc II}]$\lambda$5755 &
      [S~{\sc II}]$\lambda$6731 & [S~{\sc II}]$\lambda$6717\\
      \noalign{\smallskip} \hline \noalign{\smallskip} 250 & $1.186\times
      10^{-3}$ & $5.310\times 10^{-4}$ & $8.691\times 10^{-6}$ & $1.224\times
      10^{-4}$ & $8.799\times 10^{-5}$ \\ 500 & $1.025\times 10^{-3}$ &
      $1.581\times 10^{-4}$ & $3.069\times 10^{-6}$ & $8.937\times 10^{-5}$ &
      $6.541\times 10^{-5}$ \\ 750 & $8.896\times 10^{-4}$ & $7.773\times
      10^{-5}$ & $1.502\times 10^{-6}$ & $4.252\times 10^{-5}$ & $3.145\times
      10^{-4}$ \\ \noalign{\smallskip} \hline \end{tabular} \] \end{table*}

   \begin{table*} \caption{Peak line ratios to H$\alpha$} \label{ratios} \[
      \begin{tabular}{lllll} \hline \noalign{\smallskip} Time (yrs) & [N~{\sc
      II}]$\lambda$6583 & [N~{\sc II}]$\lambda$5755 & [S~{\sc
      II}]$\lambda$6731 & [S~{\sc II}]$\lambda$6717\\ \noalign{\smallskip}
      \hline \noalign{\smallskip} 250 & 2.000 & 0.0442 & 0.636 & 0.454\\ 500
      & 1.498 & 0.0308 & 0.844 & 0.559\\ 750 & 2.047 & 0.0436 & 1.112 &
      0.770\\ \noalign{\smallskip} \hline \end{tabular} \] \end{table*}


\begin{figure}
\psfig{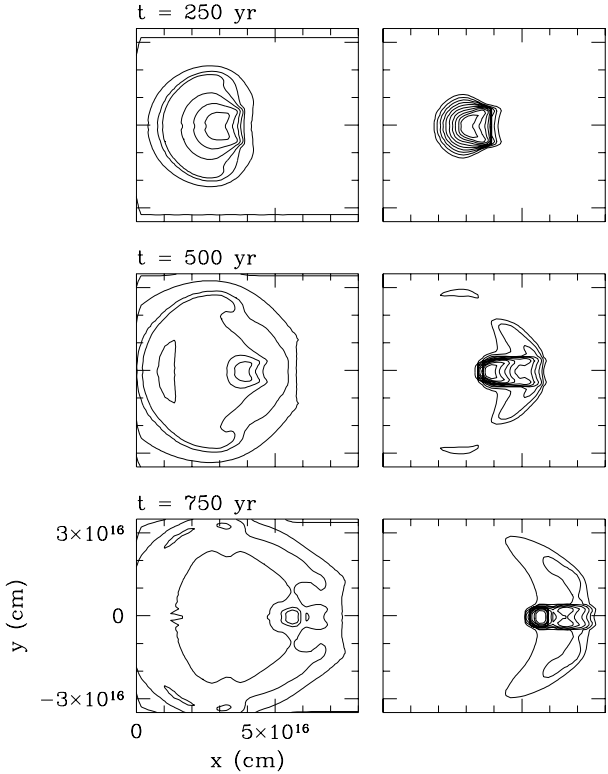}
\vskip 0.1cm
\caption{H$\alpha$ and [N {\sc II}] images for times 250, 500 and 750
years. The contours span a dynamic range of 1000 in steps of a factor 2}
\label{images}%
\end{figure}

\section{Comparison with observations}

The number of ions included in our numerical simulation makes it relatively
straightforward to produce synthesized observations of various kinds which
can be compared to the observed FLIERs. In Table~1 we show some line
strengths derived from the snapshots at $t=250$, 500, and 750~years. The top
part lists the total luminosities (in solar luminosities) emitted by the
clump material. We limit ourselves to the clump material to exclude the large
amount of H$\alpha$ emission produced by the environment. The lower part of
the table shows the maximum ratio of two forbidden lines of [N~{\sc II}] and
[S~{\sc II}] to H$\alpha$, derived using the whole image, in this case
including the environmental H$\alpha$ emission. However, the resulting ratios
are not very different if one excludes the contribution from the
environment. The model reproduces the main observational characteristic of
FLIERs and ansae, namely that they are very bright in [N~{\sc II}]$\lambda
6583$. Note that we use `normal' nitrogen abundances in the model and
therefore support the view of \cite{Dopita97} and \cite{AlexBal97} that the
high [N~{\sc II}]/H$\alpha$ ratios do not necessarily imply high nitrogen
abundances, but can be produced in these type of ionization fronts.

The other line strengths are also not far off from the ones found by
\cite{Balickea94}~(1994). Using the ratio of the red [S~{\sc II}] lines one
can derive an average electron density, which turns out to be around
1500~cm$^{-3}$ at all epochs. Similarly the ratio of the two [N~{\sc II}]
lines gives an electron temperature, which is around $10^4$~K. The density
value matches the observed one very well, but the temperature appears to be
slightly higher than the observed one (cf.~\cite{Balickea94}~1994).

\begin{figure}
\psfig{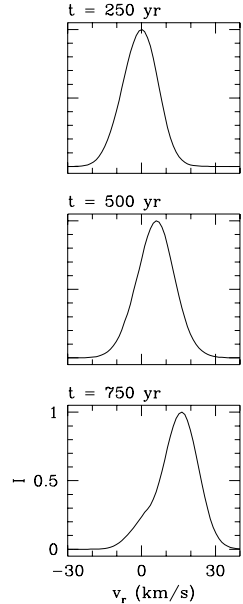}
\vskip 0.1cm
\caption{Line profiles for [N~{\sc II}]$\lambda 6583$ at times 250, 500 and
750 years. The inclination of the flow direction with respect to the observer
is taken as $45^\circ$ and the instrumental broadening is 7~km~s$^{-1}$}
\label{profiles}%
\end{figure}

Figure~\ref{images} shows images of the plasmons in the H$\alpha$ and [N~{\sc
II}]$\lambda 6583$ lines projected at an angle of $0^\circ$ in the sky. The
typical morphology is that of a little arc pointing away from the star, both
in H$\alpha$ and [N~{\sc II}]. However, in H$\alpha$ the contrast between the
plasmon and the environment is much lower than in the [N~{\sc II}] line,
making the object less noticeable, which is in accord with the observations
of FLIERs and the other low-ionization knots. [O~{\sc III}]$\lambda 5007$
(not shown) is similar to H$\alpha$, again as observed. The morphology
matches less well. The model images show arcs pointing away from the star,
whereas in the {\it HST}\/ observations the FLIERs appear to have mostly arcs
pointing towards the star, although there are also some instances for which
the reverse appears to be the case. Also, the cones pointing back towards the
star as seen in NGC~6826 and NGC~7009, have a reverse orientation compared to
the cones seen in our H$\alpha$ images. However, the {\it HST}\/ images also
show clearly that the real FLIERs are not homogeneous knots such as in our
model, making the comparison difficult.

Finally we show in Fig.~\ref{profiles} the profiles of the [N {\sc
II}]$\lambda 6583$ line for the three epochs that we are considering. The
instrumental broadening is assumed to be 7~km~s$^{-1}$ and the objects are
moving away from us at an angle of $45^\circ$. At $t=250$~years the
instrumental broadening is enough to hide any motion of the plasmon, but at
the two later times one sees how the line becomes shifted due to the motion
of the plasmon. At $t=500$~years this shift is about 7~km~s$^{-1}$, at
$t=750$~years, 17~km~s$^{-1}$. At $t=750$~years the line profile clearly
deviates from thermal, showing a low velocity tail. The observed line
profiles in [N~{\sc II}] are typically 30~km~s$^{-1}$ wide, perhaps due to
the fact that we see an ensemble of evaporating clouds.

Assuming that the clump is initially moving with the same velocity as the
environment, this environmental velocity should be added to obtain the space
velocities of the clump. The value of the environmental velocity depends on
the kinematics of the nebula, but for elliptical PNe can be estimated to lie
between 10 and 50~km~s$^{-1}$, which gives absolute FLIER velocities in the
range 30--70~km~s$^{-1}$, close to the observed values. As we saw in Sect.~5,
the average velocity of the plasmon is about 26~km~s$^{-1}$. This average
velocity is achieved when $2.8\times 10^{-6}$~M$_\odot$ (12.5\% of the
original mass) is left. Higher velocities are achieved later on, but by this
time only a tiny fraction of the original mass is left. The plasmon reaches a
velocity of 100~km~s$^{-1}$ when only 1\% of the original mass remains. One
would need to start with very massive clumps ($\sim 10^{-4}\msun$) in order
to explain FLIERs whose velocity differs by more than 30~km~s$^{-1}$ from
their immediate environment.  For those an additional acceleration mechanism
is needed, for example a high initial clump velocity or a fast wind flowing
past the clump. We intend to explore models of these types of flow in future
papers.


\section{Conclusions}

We studied the evolution of a dense clump in the radiation field of a hot
central star of a PN using both a simple analytic approach and a detailed
numerical model. We compared the results to each other and to the
observations of FLIERs. The main conclusions from this are:
\begin{enumerate}
\item The evolution of an accelerating, photo-evaporating spherical clump can
be described by a relatively simple analytic model which gives both the
velocity, the mass and the shape of the clump with time.
\item When comparing this solution with a detailed numerical model we find an
excellent agreement, implying that the simple analytic model does indeed give
a good description of the evolution of such a photo-evaporating clump.
\item The synthesized observations compiled from the numerical model can
reproduce many features of FLIERs, such as typical line strengths (especially
the strong [N~{\sc II}] lines), the average velocity difference with the
environment, and perhaps the morphology. The reproduction of strong [N~{\sc
II}] lines is noteworthy since we use standard abundances.  We therefore
conclude that the photo-evaporating clump model is moderately successful in
explaining the FLIER phenomenon. However, some of the morphological features
seen in the {\it HST}\/ observations definitely need further studying and
modelling.
\end{enumerate}

The model we studied here does not address the origin of the clump. Given
that the FLIERs lie along the major axis of the PN, it is logical to assume
that their origin is related to the cause of the asphericity of the
nebula. There are then two options, either the clump was formed very near the
star, perhaps through some process involving a disk or a jet (see
\cite{Soker96} and references therein), or during the early phases of the
interaction between the stellar fast wind and the surrounding material
(\cite{FrBaLi96}). The first explanation received some support from the
reported high nitrogen abundances, but as was shown here, as well as in
\cite{Dopita97} and \cite{AlexBal97}, the nitrogen abundance does not need to
be exceptional if one assumes the FLIERs to contain an ionization
front. Recent {\it NICMOS}\/ results on the proto-PN CRL~2688 seem to
indicate that dense knots indeed form during the proto-PN phase\footnote{See
http://oposite.stsci.edu/pubinfo/PR/97/11.html}.

The analytic solution we derived applies to any photo-evaporating clump,
and could therefore be applied to other cases such as the cometary globules
found in the Helix nebula (\cite{ODellHan96}; \cite{Meaburnea97}), and dense
knots near massive stars in star formation regions.


\begin{acknowledgements}
GM thanks the Instituto de Astronom{\'\i}a for a pleasant stay during which most
of the work in this paper was done. AR and JC acknowledge support from
CONACYT, DGAPA (UNAM) and the UNAM/Cray research program. GM and PL are both
supported by the Swedish Natural Science Research Council (NFR).  AN--C's
research is supported by NASA LTSA program with the Jet Propulsion Laboratory
of the California Institute of Technology.  WS acknowledges the receipt of a
PPARC associateship.
\end{acknowledgements}



\end{document}